\documentclass[seceq,preprint]{ptptex}

\usepackage{amssymb}
\usepackage{here}
\usepackage[dvips]{graphicx}

\notypesetlogo
\preprintnumber{HUPD0508}

\title{
Influence of QED Corrections on the Orientation of Chiral
Symmetry Breaking in the NJL model
}

\author{
 {\scshape Takahiro Fujihara,$^1$ Tomohiro Inagaki$^2$} and 
 {\scshape Daiji Kimura$^3$}
}

\inst{
 $^{1,3}$Department of Physics, Hiroshima University,
 Higashi-Hiroshima, \\
 739-8526, Japan \\
 $^2$Information Media Center, Hiroshima University,
 Higashi-Hiroshima, \\
 739-8521, Japan
}
\abst{
We study QED corrections to chiral symmetry breaking 
in the Nambu--Jona-Lasinio (NJL) model with two flavors of quarks.
In this model, the isospin symmetry is broken by the differences between the current 
quark masses and the electromagnetic charges of the up and down 
quarks.
To leading order in the $1/N$ expansion, we calculate 
the effective potential of the model with one-loop QED corrections
at finite temperature.
 Evaluating the effective potential,
we study the influence of the isospin symmetry breaking on the 
orientation of chiral symmetry breaking.
 The current quark mass plays an essential role in maintaining
the orientation of the chiral symmetry breaking.
 If the average of the up and down quark masses is small enough,
we find a phase in which the pion field has non-vanishing
expectation value and dynamical {\it CP} violation takes place. 
}

\begin{document}
\maketitle

\section{Introduction}

Quantum chromodynamics (QCD) is the first principle in describing
the physics of quarks and gluons.
 The QCD Lagrangian has the global flavor symmetry
{\it SU}$_{\rm L}(N_{\rm f}) \otimes$ {\it SU}$_{\rm R}(N_{\rm f})$
for the $N_{\rm f}$ flavors of massless quarks.
 In nature, quarks have different current masses
for all flavors, and the global flavor symmetry
is an approximate symmetry for light quarks.
Owing to asymptotic freedom in the QCD interaction, the coupling 
constant grows at low energy scales, and the hadronic phase is 
realized.
 In the hadron phase, the approximate chiral symmetry
is dynamically broken.

The broken chiral symmetry is restored in certain extreme 
environments, e.g. high temperature and high density.
This restoration of chiral symmetry is a non-perturbative phenomenon in QCD.
The phase structure of QCD has been studied in some low energy
effective theories\cite{Namb, Klev, Hats_94} and through numerical 
calculations in lattice QCD. From these results,
it is conjectured that the chiral symmetry should be restored
near the critical temperature, $T \lesssim 200$ MeV.
Recently, an experimental study at RHIC found evidence 
of symmetry restoration from hadronic matter to a state of 
deconfined partonic matter\cite{BRAH}.

The effect of an electromagnetic field on chiral symmetry breaking
is complicated. 
Low energy effective theories provide possible approaches 
for studying non-perturbative QCD phenomena in an external 
electromagnetic field. The NJL model is the simplest model 
in which the chiral symmetry is broken dynamically.
The dynamical origin of symmetry breaking in the NJL model has been 
studied in the case that there exists an external electromagnetic 
field \cite{KSI}.
The combined effect of both an external electromagnetic field and 
 gravity has also been investigated in 
Refs. \citen{Inagaki:1997nv} and \citen{Elizalde:1997aw}.
Then it was found that the external electromagnetic fields can induce 
a very wide variety of phases.

Radiative QED corrections play an important role in the 
physics of quarks and hadrons, even if we consider a system 
with no external electromagnetic field.
The difference between the electromagnetic 
charges of up and down quarks breaks the {\it SU}(2) isospin symmetry. 
The difference between the masses of up and down quarks also breaks the isospin 
symmetry. Because of the isospin breaking, it is conjectured that a variety 
of phases are realized inside quark matter through QED 
corrections. If the sum of the up and down quark masses, 
$m_{\rm u}+m_{\rm d}$, is small enough, there is a possibility
that a pion field will develop a non-vanishing vacuum expectation
value\cite{Creu}.

We assume that the NJL model is a phenomenological
low energy effective theory of QCD and that the fundamental
QCD Lagrangian is invariant under the {\it CP} transformation.
Thus, the quark mass should be real and there should be no $\theta$-term.
Because of the anomaly, the chiral {\it U}(1) transformation, which flips
the sign of the quark mass, produces the $\theta$-term in QCD.
This is only a phase convention for quark fields, and
such a $\theta$-term does not induce any {\it CP} violating phenomena.
The situation may be different after dynamical chiral symmetry
breaking. The orientation of the vacuum is fixed, and thus the {\it CP}
symmetry can also be broken.
This is an example of the dynamical {\it CP} violation proposed by 
Dashen\cite{Dash_71}.
To observe the phase structure of the theory, it seems more
convenient to choose a phase convention for the quark fields in which the
Lagrangian has no $\theta$-term. There is no degree of freedom that
allows us to set the quark mass to a positive value in this convention.
For this reason,
both positive and negative quark masses should be considered.

In the present paper, we study the influence of the current quark mass
and the quantum corrections of the electromagnetic fields on the chiral 
symmetry breaking in the NJL model. The order parameters of the chiral 
symmetry breaking are given by the vacuum expectation values of the
$\sigma$ and $\pi^a$ fields. These expectation values are found
by determining the minimum of the effective potential.
 First, we calculate the effective potential of the NJL model
to leading order in the $1/N$ expansion.
 Evaluating it, we derive the phase structure of the NJL model.
 Next, the gauged NJL model is considered. We calculate 
the effective potential with one-loop QED corrections and evaluate the 
phase structure of the gauged NJL model. The charged and neutral pion 
mass difference is investigated in \S 4. As is well-known, a higher-order 
correction in the $1/N$ expansion is essential to elucidate the pion mass 
difference. We introduce it through phenomenological meson kinetic 
terms. The phase structure is evaluated in the gauged NJL model with 
additional meson kinetic terms.
In \S 5 we study the temperature effect in the imaginary time 
formalism. The behavior of the effective potential is evaluated in both 
the NJL model and the gauged NJL model near the critical temperature, 
$T\sim 170$ MeV.
In the case of small $|m_{\rm u}+m_{\rm d}|$, there is a parameter
space in which the pion field acquires a non-vanishing vacuum expectation
value. In \S 6, the dynamical origin of {\it CP} violation is studied
within the NJL model for small $|m_{\rm u}+m_{\rm d}|$.
 Finally, we give some concluding remarks.

\section{NJL model with finite current quark mass}
We start from the Lagrangian density of the two-flavor NJL model with 
finite current quark mass. It is defined by
\begin{eqnarray}
{\cal L}_{\rm m} = \bar\psi (i \gamma^\mu \partial_\mu -M)\psi
           + \frac{G}{2 N} \left[ (\bar\psi \psi)^2 
           + (\bar\psi i\gamma_5\tau^a \psi)^2 \right] ,
\label{lag} 
\end{eqnarray}
where $G$ is the coupling constant of the four-fermion interaction,
$N$ represents the number of colors, $M$ is a $2\times 2$ matrix
which contains the current quark masses for the up and down quarks, 
and $\tau^a \ (a=1,2,3)$ are the Pauli matrices of the isospin vector.
We assume a diagonal form for the mass matrix, 
$M \equiv {\rm diag} (m_{\rm u},\ m_{\rm d})$.
In our phase convention it cannot be assumed that the mass matrix is positive.
In four dimensions this model is not renormalizable, 
because the coupling constant $G$ scales with the mass as
$(\mbox{mass})^{-2}$. 
The model is defined with a regularization parameter and is
regarded as a low energy effective theory of QCD.

In the massless limit, $M \rightarrow 0$, the Lagrangian density 
(\ref{lag}) is invariant under the isospin symmetry $SU_{\rm V}(2)$ 
transformation,
$\psi(x) \to {\rm exp} (i \theta^a \tau^a / 2) \psi(x)$,
and the chiral transformation, 
$\psi(x) \to {\rm exp} (i \gamma_5 \theta_5^a \tau^a / 2) \psi(x)$.
In this case, the chiral charge is defined by
$Q_5^a = \int d^3 x \bar{\psi} \gamma_0 \gamma_5 (\tau^a/2) \psi$.
The commutation relations between the chiral charge and  fermion bi-linears
are given for the composite scalar operator $\bar\psi \psi$ as
\begin{eqnarray}
\left[i Q_5^a,\ \bar\psi \psi(x) \right] 
&=& \bar\psi i \gamma_5 \tau^a \psi(x)
\label{comm_sc}
\end{eqnarray}
and for the composite pseudo-scalar operator 
$\bar\psi i\gamma_5\tau^a \psi$ as
\begin{eqnarray}
\left[i Q_5^a,\ \bar\psi i\gamma_5\tau^b \psi(x) \right] 
&=& - \delta^{ab} \bar\psi \psi(x) .
\label{comm_ps}
\end{eqnarray}
If the composite operators $\bar\psi i \gamma_5 \tau^a \psi$ and/or 
$\bar\psi \psi$ develop non-vanishing expectation values, 
the chiral symmetry is dynamically broken.

In the real world, up and down quarks have small current masses.
The mass term is not invariant under the chiral transformation; 
it breaks the chiral symmetry explicitly.
In the case $m_{\rm u} \not= m_{\rm d}$, the isospin symmetry,
$SU_{\rm V}(2)$, is also broken.

For simplicity, we rewrite the Lagrangian density (\ref{lag})
using the auxiliary fields method. First, we introduce the auxiliary scalar
field $\sigma \simeq - (G / N) \bar\psi \psi$ and pseudo-scalar
fields $\pi^a \simeq - (G / N) \bar\psi i \gamma_5 \tau^a \psi$.
Thus the Lagrangian density (\ref{lag}) is rewritten as
\begin{eqnarray}
{\cal L}_{\rm m} = \bar\psi (i \gamma^\mu \partial_\mu - M 
    - \sigma - i \gamma_5 \tau^a \pi^a ) \psi
    - \frac{N}{2 G} \left[ \sigma^2 + (\pi^a)^2 \right] .
\label{lag_m_au}
\end{eqnarray}
A neutral pion $\pi^0$ and charged pion fields $\pi^{\pm}$ 
are defined by
\begin{eqnarray}
  \left(
    \begin{array}{cc}
      \pi^0 & \sqrt{2} \pi^+ \\
      \sqrt{2} \pi^- & - \pi^0
    \end{array}   
  \right) 
\equiv
\tau^a \pi^a . 
\label{pi_pm}
\end{eqnarray}

To study the phase structure of the model, we evaluate the effective
potential. The ground state is found by finding the minimum of the
effective potential. Here, we start with the following generating functional of Green 
functions:
\begin{eqnarray}
Z_{\rm m} &\equiv& \int {\cal D} \psi {\cal D} \bar\psi 
     {\cal D} \sigma {\cal D} \pi \exp \left( i \int d^4 x
     {\cal L}_{\rm m} \right)
\nonumber \\
  &=& \int {\cal D} \sigma {\cal D} \pi \exp 
     \left[i N \left\{ -\frac{1}{2 G} \int d^4 x
     \left[ \sigma^2 + (\pi^a)^2 \right] \right.\right. 
\nonumber \\ 
  &&  - i \ln \det (i \gamma^\mu \partial_\mu 
      - M - \sigma - i \gamma_5 \tau^a \pi^a )
     \biggr\} \biggr] .
\label{z_m}
\end{eqnarray}
Owing to translational invariance, the expectation value 
of the auxiliary fields should be constant. 
Furthermore, the quantum corrections from the internal lines of the
auxiliary fields disappear in the large $N$ limit. 
To leading order in the $1/N$ expansion, the path integral 
over the auxiliary fields in Eq. (\ref{z_m}) gives only an 
overall factor. Thus the effective potential is found to be
\begin{eqnarray}
V_{\rm m}(\sigma,\pi^a)
&\equiv& - \frac{1}{2iN \int d^4 x} \ln Z_{\rm m} 
\nonumber \\
&=& \frac{1}{4 G} 
     \left[ \sigma^2 + (\pi^a)^2 \right]
    + \frac{i}{2 \int d^4 x} \ln \det (i \gamma^\mu \partial_\mu 
      - M - \sigma - i \gamma_5 \tau^a \pi^a ) .
\label{epot}
\end{eqnarray}

The current quark masses for up and
down quarks are much smaller than the $\pi$ meson scale.
 For this reason, we evaluate the 
effective potential up to second 
order in the current quark mass and apply the expansion
\begin{eqnarray}
&& i \ln \det ( i \gamma^\mu \partial_\mu   
       - M - \sigma - i \gamma_5 \tau^a \pi^a )
\nonumber \\
&&
  = i {\rm tr} \ln (i \gamma^\mu \partial_\mu 
       - \sigma - i \gamma_5 \tau^a \pi^a) 
     + {\rm tr} \sum_{n=1}^\infty I_n
\nonumber \\
&&  = i {\rm tr} \ln (i \gamma^\mu \partial_\mu 
       - \sigma - i \gamma_5 \tau^a \pi^a) 
      + {\rm tr} (I_1+I_2) + O(m^3) ,
\label{exp_m}
\end{eqnarray}
where tr represents the trace over the flavor, spinor and
space-time coordinates, and $I_n$ is given by
\begin{eqnarray}
I_n \equiv \frac{1}{i n}
  \left( \frac{M}{i \gamma^\mu \partial_\mu 
  - \sigma - i \gamma_5 \tau^a \pi^a + i \varepsilon} \right)^n .
\label{i_n}
\end{eqnarray}

In four dimensions, Eq. (\ref{exp_m}) is divergent.
To obtain a finite result, we introduce the three-momentum cutoff 
$\Lambda_{\rm f}$.\footnote{We use a three-momentum cutoff 
to obtain a finite result, employing the same regularization method
even at finite temperature.} 
After some straightforward  calculations, a finite expression for 
the effective potential is obtained:
\begin{eqnarray}
V_{\rm m}(\sigma,\pi^a)
&=& \frac{1}{4G} \left[ \sigma^2 + (\pi^a)^2 \right]
  - \frac{1}{8\pi^2} f(\sigma'^2; \Lambda_{\rm f}{}^2)
\nonumber \\
&& 
  - \frac{1}{4\pi^2} (m_{\rm u} + m_{\rm d}) \sigma
  g(\sigma'^2; \Lambda_{\rm f}{}^2)
\nonumber \\
&&  
  - \frac{1}{8\pi^2} (m_{\rm u}{}^2 + m_{\rm d}{}^2)
  g(\sigma'^2; \Lambda_{\rm f}{}^2)
\nonumber \\
&& 
  - \frac{1}{4\pi^2} 
  \left[ (m_{\rm u}{}^2 + m_{\rm d}{}^2) \sigma^2
  + (m_{\rm u} - m_{\rm d})^2 \pi^+ \pi^- \right]
\nonumber \\
&&
  \times \left( \frac{\Lambda_{\rm f}}
    {\sqrt{\Lambda_{\rm f}{}^2 + \sigma'^2}}
  - \ln \frac{\Lambda_{\rm f} + 
      \sqrt{\Lambda_{\rm f}{}^2 + \sigma'^2}}
    {\sqrt{\sigma'^2}} \right) , 
\label{v_m} 
\end{eqnarray}
where we have defined $\sigma'^2 \equiv \sigma^2 + (\pi^a)^2$, and the functions
$f(s^2; t^2)$ and $g(s^2; t^2)$ are given by
\begin{eqnarray}
f(s^2; t^2) 
&\equiv& (2t^2 + s^2) \sqrt{t^2(t^2 + s^2)} 
 - s^4 \ln \frac{\sqrt{t^2} + \sqrt{t^2 + s^2}}{\sqrt{s^2}} \ ,
\label{fst}
\\
g(s^2; t^2) &\equiv& \sqrt{t^2(t^2 + s^2)} 
 - s^2 \ln \frac{\sqrt{t^2} + \sqrt{t^2 + s^2}}{\sqrt{s^2}} \ .
\end{eqnarray}

The effective potential (\ref{v_m}) is symmetric with respect to 
the isospin transformation up to terms linear in the
current quark mass.
Because the isospin breaking contribution appears only at orders
higher than $O(m^2)$, it gives only a small effect.
For $m_u=-m_d \ne 0$ the  terms of $O(m)$ in Eq.~(2.10) vanish and the
 terms of $O(m^2)$ mainly contribute to the orientation of the chiral 
symmetry breaking. It induces the pion condensation.
To determine the contribution from the isospin breaking, we evaluate
the behavior of the  effective potential (\ref{v_m})  
as the difference between the masses of the up and down quarks varies.
In Table \ref{lis_mas}, we list the values of $\sigma$ and $\pi$ at the 
local minimum of the effective potential in the $\pi^a=0$ plane,
$V_{\rm m}(\sigma, \pi^a=0)$, and the $\sigma=0$ plane, 
$V_{\rm m}(\sigma=0, \pi^a)$, respectively. We also list
the values of the effective potential for both points.
The vacuum state corresponds to the point with the 
smallest value of the effective potential. In our model,
this state is located in the $\sigma=0$ or $\pi^a=0$ plane. 

As seen in Table \ref{lis_mas}, the quark mass mainly contributes to
the $\sigma$ direction of the effective potential,
while the isospin breaking in the quark mass term induces
the $\langle \pi^{1,2} \rangle_{\sigma=0,\pi^3=0} 
- \langle \pi^3 \rangle_{\sigma=0,\pi^{1,2}=0}$ difference.
However, the ratio of the depths of the effective potential,
\begin{equation}
r_{\rm V} \equiv
\left| \frac
  {V(\sigma=0, \langle \pi^{1,2} \rangle_{\sigma=0, \pi^3=0}, \pi^3=0)
 - V(\sigma=0, \pi^{1,2}=0, \langle \pi^{3} \rangle_{\sigma=0, \pi^{1,2}=0})}
  {V(\sigma=0, \langle \pi^{1,2} \rangle_{\sigma=0, \pi^3=0}, \pi^3=0)
 + V(\sigma=0, \pi^{1,2}=0, \langle \pi^{3} \rangle_{\sigma=0, \pi^{1,2}=0})}
\right| ,
\label{rv}
\end{equation}
is approximately $0.008\%$ for $m_{\rm u}=2.5$ MeV and $m_{\rm d}=6.5$ MeV.

\begin{table}[t]
\begin{center}
\caption{Influence of the isospin breaking on the effective potential.
The columns labeled
$\langle\sigma\rangle_{\pi^a=0}$ and $\langle\pi^a\rangle_{\sigma=0}$ 
list the local minima of $V_{\rm m}(\sigma, \pi^a=0)$ and 
$V_{\rm m}(\sigma=0, \pi^a)$, respectively. 
}
\begin{tabular}{|r|r|r|r|r|r|} \hline
  {\footnotesize $m_{\rm u}$} 
 & {\footnotesize $m_{\rm d}$} 
 & {\footnotesize $\langle\sigma\rangle_{\pi^a=0}$} 
 & {\footnotesize $\langle\pi^a\rangle_{\sigma=0}$} 
 & {\footnotesize $V_{\rm m}(\langle\sigma\rangle_{\pi^a=0}, \pi^a=0)$} 
 & {\footnotesize $V_{\rm m}(\sigma=0,\langle\pi^a\rangle_{\sigma=0})$} \\
  {\footnotesize (MeV)} 
 & {\footnotesize (MeV)} 
 & {\footnotesize (GeV)} 
 & {\footnotesize (GeV)} 
 & {\footnotesize (GeV${}^4$)} 
 & {\footnotesize (GeV${}^4$)} \\ \hline
  {\footnotesize $4.5$}
 & {\footnotesize $4.5$}
 & {\footnotesize 0.2829} 
 & {\footnotesize 0.2719} 
 & {\footnotesize $-9.107 \times 10^{-5}$} 
 & {\footnotesize $-6.610 \times 10^{-5}$} \\ \hline
  {\footnotesize $2.5$}
 & {\footnotesize $6.5$}
 & {\footnotesize 0.2829} 
 & {\footnotesize 0.2718} 
 & {\footnotesize $-9.105 \times 10^{-5}$} 
 & {\footnotesize $\pi^{1,2}: -6.608 \times 10^{-5}$,} \\
 &  &  &  
 &  & {\footnotesize $\pi^3: -6.609 \times 10^{-5}$} \\ \hline
  {\footnotesize $-4.5$}
 & {\footnotesize $4.5$}
 & {\footnotesize 0.2718} 
 & {\footnotesize $\pi^{1,2} = 0.2718$,}
 & {\footnotesize $-6.604 \times 10^{-5}$} 
 & {\footnotesize $\pi^{1,2}: -6.604 \times 10^{-5}$,} \\
 &  &   
 & {\footnotesize $\pi^3 = 0.2719$}
 &  & {\footnotesize $\pi^3: -6.610 \times 10^{-5}$}\\ \hline
  {\footnotesize $-2.5$}
 & {\footnotesize $6.5$}
 & {\footnotesize 0.2769} 
 & {\footnotesize 0.2718} 
 & {\footnotesize $-7.706 \times 10^{-5}$} 
 & {\footnotesize $\pi^{1,2}: -6.603 \times 10^{-5}$,} \\
 &  &  &    
 &  & {\footnotesize $\pi^3: -6.609 \times 10^{-5}$} \\ \hline
\end{tabular}
\label{lis_mas}
\end{center}
\end{table}

\begin{figure}[t]
 \begin{center}
  \includegraphics[width=70mm,clip]{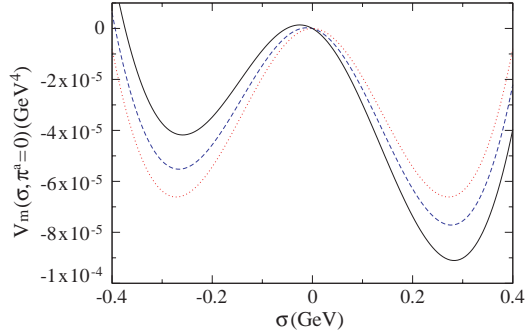}
    \caption{\small Behavior of the effective potential $V(\sigma, \pi^a=0)$ 
      for $m_{\rm u}+m_{\rm d}=0$ MeV (red curve),
  $m_{\rm u}+m_{\rm d}=4$ MeV (blue curve) 
      and $m_{\rm u}+m_{\rm d}=9$ MeV (black curve).}
  \label{sig_m}
 \end{center}
\end{figure}

We plot the behavior of the effective potential in the $\pi^a=0$ 
plane in Fig.~\ref{sig_m}.
 It is seen that increasing 
the total current quark mass of the up and down quarks, $m_{\rm u}+m_{\rm d}$, 
causes the effective potential to increase for a negative $\sigma$ and 
decrease for a positive $\sigma$. Thus the global minimum of the 
effective potential appears at a positive value of $\sigma$. 

\begin{figure}[t]
 \begin{center}
  \includegraphics[width=70mm,clip]{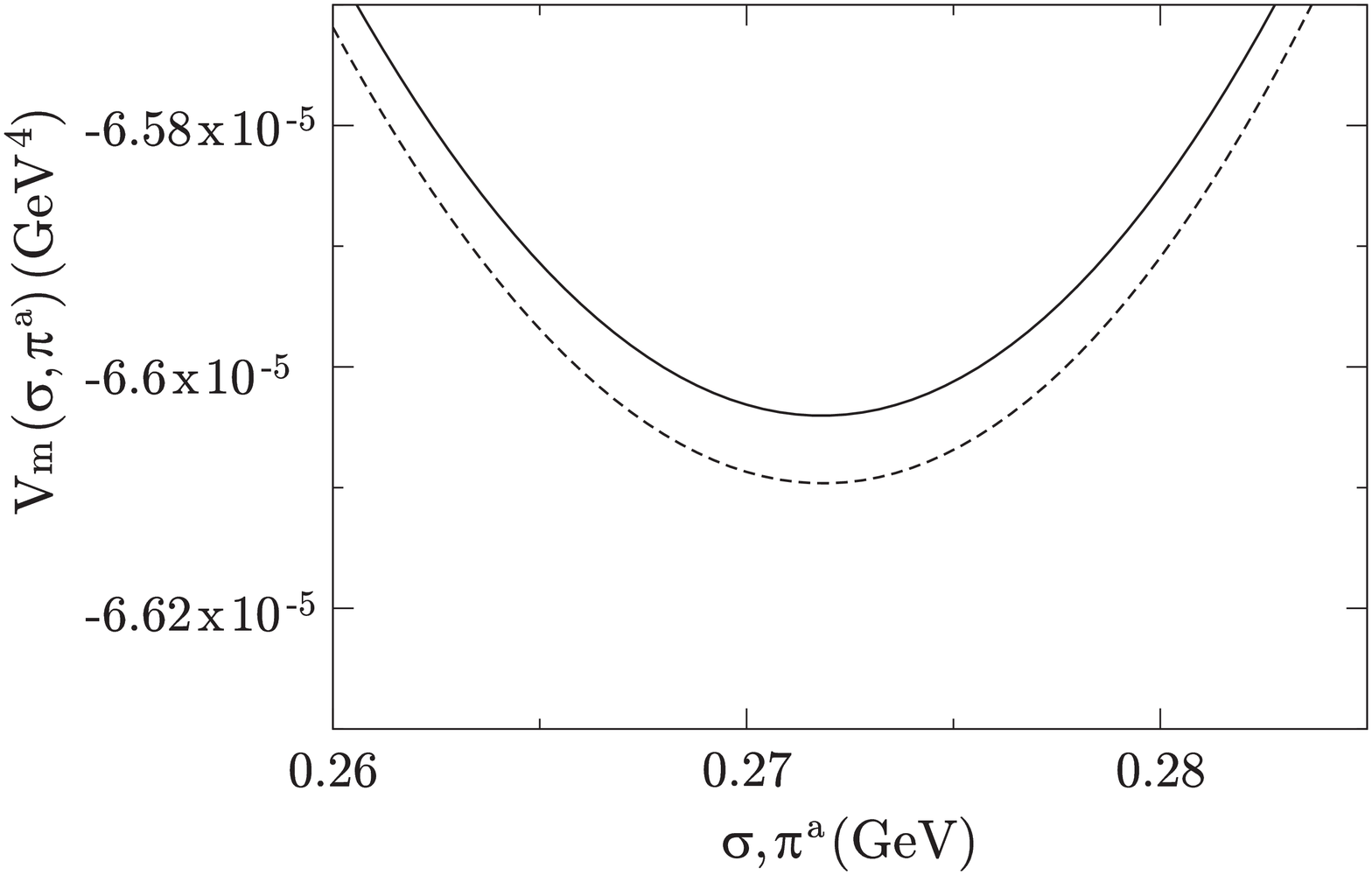}
    \caption{\small Behavior of the effective potential (\ref{v_m})
      for $m_{\rm u}+m_{\rm d}=0$ MeV in the 
      $\pi^3=0$ plane (dashed curve)
      and the $\sigma=\pi^{1,2}=0$ plane (solid curve).}
  \label{v_mcomp}
 \end{center}
\end{figure}

There is freedom in the choice of the sign of the fermion mass term.
In Fig.~\ref{v_mcomp} we plot the behavior of the effective potential
for $m_{\rm u}=-m_{\rm d}$. 
In this case, it depends on 
$\sigma^2+(\pi^1)^2+(\pi^2)^2$ and $(\pi^3)^2$.
In the figure,
we plot the effective potential in the $\pi^3=0$ plane and
the $\sigma=\pi^{1,2}=0$ plane.
It is seen that in this case, the effective potential exhibits behavior
similar to that the sigma axis, $\pi^a=0$, and the charged pion axis, $\sigma=\pi^3=0$.
As seen in 
Fig.~\ref{v_mcomp}, the global minimum of the effective potential 
lies in the $\sigma=\pi^{1,2}=0$ plane.
This implies that the neutral pion field develops a non-vanishing 
expectation value, $\langle \pi^3 \rangle \neq 0$. It is found that 
pion condensation takes place for 
$|m_{\rm u}+m_{\rm d}|\lesssim 0.020$ MeV.
Here, we consider only the leading-order in the 1/N expansion.
Higher order corrections can modify the small difference of the
lines in Fig.~\ref{v_mcomp}.

If $|m_{\rm u}|+|m_{\rm d}|$ is fixed at a realistic value ($\sim 9$ MeV)
and both $m_{\rm u}$ and $m_{\rm d}$ are positive, the global minimum of
the effective potential lies in the $\pi^a=0$ plane.
An enhancement of the symmetry breaking in the $\pi^a=0$ plane
is mainly caused by the isospin symmetric terms, which are
proportional to $|m_{\rm u}+m_{\rm d}|$ in Eq.~(\ref{v_m}).
Only $\sigma$ develops a non-vanishing expectation value.
For small values of $|m_{\rm u}+m_{\rm d}|$, the situation is different.
 The higher-order terms, specifically, {\it O}($m^2$),
in Eq.~(\ref{v_m}) break the isospin symmetry
and yield non-trivial corrections to the effective potential.
These corrections can move the global minimum of the effective
potential away from the $\pi^a=0$ plane.
 If we take $m_u\simeq -m_d$, the neutral pion develops a
non-vanishing vacuum expectation value, and the vacuum state breaks
the discrete symmetry under the parity transformation, as 
pointed out in Ref.~8).

In such calculations,
the mass scales of the effective potential are determined from
the physical observables.
For this purpose, we calculate the pion decay constant $f_\pi$ 
and the pion mass $m_\pi$ and fix the three-momentum cutoff 
$\Lambda_{\rm f}$ and the coupling constant $G$. \cite{Inag_PTP}
Here we set the current quark mass as $m_{\rm u} + m_{\rm d} = 9$ MeV
and $m \equiv (m_{\rm u} + m_{\rm d}) / 2$.
 In this case, the minimum of the effective potential is at
$\sigma = \langle \sigma \rangle$ and $\pi^a = 0$.
Taking the partial derivative of the generating functional 
(\ref{z_m}) with respect to $\pi$ twice, we obtain the two-point 
vertex function for the pion field, $\Gamma_\pi^{(2)} (p)$.
At $\sigma = \langle \sigma \rangle$ and $\pi = 0$,
it is given by
\begin{eqnarray}
\Gamma_\pi^{(2)} (p) &=&
 - \frac{N}{G} - N \int^{\Lambda_{\rm f}}
 \frac{d^4 k}{i(2\pi)^4} {\rm tr}
 \frac{-i \gamma_5 \tau^a}{\gamma^\mu k_\mu - M - \sigma}
 \frac{-i \gamma_5 \tau^b}{\gamma^\nu (k+p)_\nu - M - \sigma} 
\nonumber \\
&=& - \frac{N}{G}
  + \frac{N\delta^{ab}}{2 \pi^2}
  \Biggl[ 2\Lambda_{\rm f} \sqrt{\Lambda_{\rm f}{}^2+(\sigma+m)^2}
\nonumber \\
&& - \sqrt{p^2} \sqrt{4(\sigma+m)^2-p^2} \arctan 
   \frac{\Lambda_{\rm f} \sqrt{p^2}}
     {\sqrt{\Lambda_{\rm f}{}^2+(\sigma+m)^2} 
   \sqrt{4(\sigma+m)^2-p^2}} 
\nonumber \\
&& \left. - \left\{ 2(\sigma+m)^2 - p^2 \right\}
   \ln \frac{\Lambda_{\rm f} + 
         \sqrt{\Lambda_{\rm f}{}^2+(\sigma+m)^2}}{\sigma+m}
   \right] ,\nonumber
\\ 
&& {\rm for}\ 0 \leq p^2 < 4(\sigma+m)^2 . \label{2pf}
\end{eqnarray}

The wave function renormalization, $Z_\pi$, is defined as
$\Gamma_\pi^{(2)} (p) = Z_\pi^{-1} p^2 + O(p^4)$.
Evaluating the coefficient of $p^2$ in Eq. (\ref{2pf}), we
obtain
\begin{eqnarray}
Z_\pi^{-1} = \frac{N}{2\pi^2}
  \left[ \ln \frac{\Lambda_{\rm f} + 
               \sqrt{\Lambda_{\rm f}{}^2+(\sigma+m)^2}}{\sigma+m}
  - \frac{\Lambda_{\rm f}}
      {\sqrt{\Lambda_{\rm f}{}^2+(\sigma+m)^2}} \right] .
\label{z_pi}
\end{eqnarray}
The pion decay constant $f_\pi$ is given by 
\begin{equation}
  f_\pi = Z_\pi^{-1/2} \sigma .
\label{fpi}
\end{equation}
We also use the following model-independent relationship, which is known
as, the Gell-Mann--Oakes--Renner relation \cite{Gell}:
\begin{equation}
f_\pi{}^2 m_\pi{}^2 = \frac{N}{G} m \sigma.
\label{fpi2}
\end{equation}
Inserting the experimental values $f_\pi = 91.9$ MeV and 
$m_\pi = 135$ MeV, which are neutral pion values,
and $m = 4.5$ MeV into Eqs.~(\ref{fpi}) and 
(\ref{fpi2}) and solving the gap equation, 
$dV_{\rm m}/d\sigma=0$, we fix the cutoff scale and the coupling
constant to $\Lambda_{\rm f} = 0.697$ GeV and 
$G = 24.8 {\rm ~GeV}^{-2}$. 
These parameter values were determined without QED corrections.
 However, the QED scale is much smaller than the typical scale for pions.
It is also appropriate to use these parameter values in the gauged NJL model
considered below.

Using Eqs.~(\ref{v_m}) and (\ref{fpi}), we can calculate the mass 
difference for the charged and neutral pions,
\begin{eqnarray}
m_{\pi^{\pm}}{}^2 - m_{\pi^0}{}^2 
\equiv 2 N Z_\pi \left.
 \left[ \frac{\partial^2 V_{\rm m}(\sigma,\pi^a)}
        {\partial (\pi^{1,2})^2}         
     -  \frac{\partial^2 V_{\rm m}(\sigma,\pi^a)}
        {\partial (\pi^{3})^2} 
 \right] \right|_{\langle \sigma \rangle, \langle \pi^a \rangle},
\label{mas_dif}
\end{eqnarray} 
where we evaluate the right-hand side of this equation at
the minimum of the effective potential, $V_m(\sigma, \pi^a)$.
We list the pion mass and the mass difference in Table II.
Due to the {\it SU}(2) isospin symmetry, the charged and  neutral pions have
the same mass for $m_u=m_d$.
In the case $m_u = -2.5$ MeV and $m_d=6.5$ MeV,
the charged and neutral pion masses become smaller than in 
the positive up and down quark mass case.
For $m_{\rm u}=-4.5$ MeV and $m_{\rm d}=4.5$ MeV, 
the vacuum expectation value is in the $\pi^3$ direction.
Non-vanishing expectation value is developed for $\langle \pi^3\rangle$.
Hence the roles of $\sigma$ and $\pi^3$ are exchanged.

 Here we use the cutoff and the coupling constant for $m_u+m_d=9$ MeV.
As seen in Table \ref{lis_mpi}, the pion masses are
quite different from 135 MeV for $m_{\rm u}=-2.5$ and $-4.5$ MeV.
In these cases, Eqs. (\ref{z_pi}) and (\ref{fpi2}) are not satisfied with
$m_\pi = 135$ MeV and $f_\pi = 91.9$ MeV.
In order to reproduce realistic values for
$m_\pi$ and $f_\pi$ in the case with negative $m_u$, 
we should modify the condition (\ref{fpi2})
and use suitable values of $\Lambda_f$ and $G$.

\begin{table}[t]
\begin{center}
\caption{Influence of the isospin breaking on the 
charged and neutral pion mass. }
\begin{tabular}{|r|r|r|r|r|r|r|} \hline
  {\footnotesize $m_{\rm u}$} 
 & {\footnotesize $m_{\rm d}$} 
 & {\footnotesize Min. of $V_m(\sigma,\pi^a)$} 
 & {\footnotesize $m_{\sigma}$} 
 & {\footnotesize $m_{\pi^\pm}$} 
 & {\footnotesize $m_{\pi^0}$} 
 & {\footnotesize $m_{\pi^\pm} - m_{\pi^0}$} \\
  {\footnotesize (MeV)} 
 & {\footnotesize (MeV)} 
 & {\footnotesize (MeV)} 
 & {\footnotesize (MeV)} 
 & {\footnotesize (MeV)} 
 & {\footnotesize (MeV)} 
 & {\footnotesize (MeV)} \\ \hline
  {\footnotesize 4.5}
 & {\footnotesize 4.5}
 & {\footnotesize $\langle \sigma \rangle = 282.9$} 
 & {\footnotesize 590.4} 
 & {\footnotesize 134.0} 
 & {\footnotesize 134.0} 
 & {\footnotesize 0} \\ \hline
  {\footnotesize 2.5}
 & {\footnotesize 6.5}
 & {\footnotesize $\langle \sigma \rangle = 282.9$} 
 & {\footnotesize 590.4} 
 & {\footnotesize 134.0} 
 & {\footnotesize 133.9} 
 & {\footnotesize 0.1} \\ \hline
  {\footnotesize $-4.5$}
 & {\footnotesize $4.5$}
 & {\footnotesize $\langle \pi^0 \rangle = 271.9$} 
 & {\footnotesize 8.948}
 & {\footnotesize 8.948}
 & {\footnotesize 538.9} 
 & {\footnotesize $-530$} \\ \hline
  {\footnotesize $-2.5$}
 & {\footnotesize $6.5$}
 & {\footnotesize $\langle \sigma \rangle = 276.9$} 
 & {\footnotesize 562.4}
 & {\footnotesize 88.75} 
 & {\footnotesize 88.29} 
 & {\footnotesize 0.46} \\ \hline
\end{tabular}
\label{lis_mpi}
\end{center}
\end{table}

\section{Gauged NJL model}
Since the electric charges of the up and down quarks
are different, an electromagnetic interaction breaks 
the {\it SU}(2) isospin symmetry.
To elucidate the influence of this isospin breaking, we introduce
the QED correction. The QED interaction causes the derivative 
in the Lagrangian density (\ref{lag}) to be replaced with a covariant one. 
Thus, the Lagrangian density is extended to that of the gauged NJL model,%
\footnote{As discussed in Ref.~\citen{Mira_89},
this model has a large anomalous dimension, and
the coupling constant is walking (not running).
Hence, it may be possible to renormalize the theory
even in four dimensions.}
\begin{equation}
{\cal L}_{\rm f} = \bar\psi \left[i \gamma^\mu 
           (\partial_\mu + ieQ A_\mu) - M \right] \psi
           + \frac{G}{2 N} \left[ (\bar\psi \psi)^2 
           + ( \bar\psi i\gamma_5\tau^a \psi)^2 \right] ,
\label{lag_f}
\end{equation}
where $Q$ represents the charge of quark fields, $Q \equiv {\rm diag}(2/3,\ -1/3)$.
In this Lagrangian density, the isospin symmetry is broken, 
except along the $\tau^3$ direction.
Since the QED corrections come from the internal photon lines at 
lowest order, we should also consider the free Lagrangian for photons,
\begin{equation}
{\cal L}_{\rm ph} = - \frac{1}{4} F_{\mu \nu} F^{\mu \nu} 
            - \frac{1}{2 \xi} (\partial_\mu A^\mu)^2 ,
\label{lag_ph}
\end{equation}
and evaluate the theory with the Lagrangian
\begin{equation}
{\cal L} = {\cal L}_{\rm f} + {\cal L}_{\rm ph} .
\label{lag_0} 
\end{equation}

\begin{figure}[t] 
 \begin{center}
  \includegraphics[width=100mm,clip]{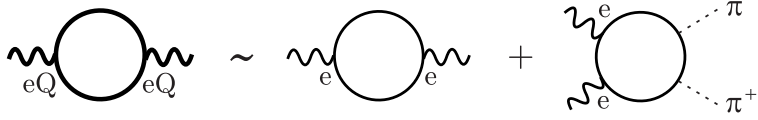}
    \caption{ \small{The graphs expressing Eq.~(\ref{i_2e}).
      The circles represent the quark propagators with the
      constituent quark mass, and the wavy lines represent photons.}
    }
  \label{pho_sel}
 \end{center}
\end{figure}

First, we evaluate the path integral for the quark fields.
Employing the auxiliary field method applied in the previous 
section, we introduce the fields $\sigma$ and $\pi$ and
calculate the generating functional. Then, replacing the derivative 
in Eq.~(\ref{z_m}) by a covariant one, we obtain
\begin{eqnarray}
Z
&=& \int {\cal D} \sigma {\cal D} \pi {\cal D} A \exp 
     \left[i\int d^4 x {\cal L}_{\rm ph}+i N \left\{ -\frac{1}{2 G} \int d^4 x 
     \left[ \sigma^2 + (\pi^a)^2 \right] \right.\right. 
\nonumber \\ 
  &&  - i \ln \det \left[ i \gamma^\mu (\partial_\mu + ieQ A_\mu)
      - M - \sigma - i \gamma_5 \tau^a \pi^a \right] 
     \biggr\} \biggr] .
\label{z_i}
\end{eqnarray}
The second line in Eq.~(\ref{z_i}) is expanded as
\begin{eqnarray}
&& i \ln \det \left[ i \gamma^\mu (\partial_\mu + ieQ A_\mu) 
       - M - \sigma - i \gamma_5 \tau^a \pi^a \right]
\nonumber \\
&&  
  = i {\rm tr} \ln (i \gamma^\mu \partial_\mu 
       - \sigma - i \gamma_5 \tau^a \pi^a) 
    + {\rm tr} \sum_{n=1}^\infty J_n ,
\label{exp_A}
\end{eqnarray}
where $J_n$ is given by
\begin{eqnarray}
J_n \equiv \frac{1}{i n}
    \left( \frac{M + eQ \gamma^\mu A_\mu}{i \gamma^\mu \partial_\mu 
    - \sigma - i \gamma_5 \tau^a \pi^a + i\varepsilon}
    \right)^n .
\label{j_n}
\end{eqnarray}
Because the electric charge, $e$, and the current quark mass, $M$,
are sufficiently small,
we evaluate Eq.~(\ref{j_n}) up to $n = 2$. At this order,
the current quark mass and photon dependent parts are separate.
The quark mass dependent part is identical to that obtained in 
the previous section. 
We calculate the photon dependent part using dimensional 
regularization and get 
\begin{eqnarray}
{\rm tr} J_{2}^{\rm ph} &=& \frac{1}{2} \int d^4 x
  \int \frac{d^4 p}{(2 \pi)^4} 
  \frac{d^D k}{i(2 \pi)^D} {\rm tr}
  \frac{1}{\gamma_\rho k^\rho - \sigma - i \gamma_5 \tau^a \pi^a}
    e_0 \mu^\epsilon Q \gamma_\mu A^\mu (p) 
\nonumber \\
&& \times
   \frac{1}{\gamma_\sigma (k^\sigma + p^\sigma) - \sigma
     - i \gamma_5 \tau^a \pi^a}
     e_0 \mu^\epsilon Q \gamma_\nu A^\nu (-p) 
\nonumber \\
&=& - \frac{1}{2} \int d^4 x \int \frac{d^4 p}{(2 \pi)^4}
   A_\mu (p) A_\nu (-p) \left[ ( 
   p^\mu p^\nu - g^{\mu \nu} p^2 )
   \Pi (p^2) \right.
\nonumber \\
&& \left. + g^{\mu\nu} \pi^+ \pi^- \Pi_{\pi\pi} (p^2) \right] .
\label{i_2e}
\end{eqnarray}
Note that there is no term proportional to $A(p)^2 \pi^0\pi^0$ in 
Eq.~(\ref{i_2e}).
 The vacuum self-energies for the photon field, $\Pi(p^2)$ and 
$\Pi_{\pi\pi}(p^2)$, are given by the diagrams in Fig.~\ref{pho_sel}. 
Performing the momentum integration for the internal fermion lines, we get
\begin{eqnarray}
\Pi (p^2) &\equiv&  \frac{5 \alpha_0}{27 \pi}  
  \left[ \frac{1}{\epsilon} - \gamma +
  \ln \frac{4 \pi \mu^2}{\sigma'^2} 
  + \frac{8}{3} - h^2
  - \frac{h}{2}(3-h^2) 
  \ln \frac{h+1}{h-1} \right] ,   
\label{pi} \\
\Pi_{\pi\pi} (p^2) &\equiv& \frac{2 \alpha_0}{\pi}
  \left(\frac{1}{\epsilon}
  - \gamma + \ln \frac{4 \pi \mu^2}{\sigma'^2} + 2
  - h \ln \frac{h+1}{h-1} \right) ,
\label{pi_p}
\end{eqnarray}
where the regularization parameter $\epsilon$ is given by
$\epsilon=(4-D)/2$, and we have defined 
$e_0 \mu^\epsilon \equiv e,\ \alpha_0 \equiv e_0{}^2 / (4\pi)$
and $h \equiv \sqrt{1 - 4 \sigma'^2 / (p^2 + i \varepsilon)}$.

These self-energies are divergent in the four-dimensional limit.
To obtain a finite result,
we must renormalize the photon self-energy. 
Here we introduce the following counter-terms into Eqs.~(\ref{i_2e})--(\ref{pi_p}):
\begin{eqnarray}
\Delta {\cal L} \equiv - \frac{1}{4} (Z_3 - 1) F_{\mu\nu} F^{\mu\nu}
 + Z_{\pi\pi} (\partial_\mu + ie A_\mu) \pi^+ 
   (\partial^\mu - ie A^\mu) \pi^-  ,
\label{del_lag}
\end{eqnarray}
where the wave function renormalizations $Z_3$ and $Z_{\pi\pi}$ are defined by
$Z_3 - 1 = - N \Pi(0)$ and 
$Z_{\pi\pi} = - (N / 2) \Pi_{\pi\pi}(0)$.
We shift the pion fields as
$\pi^\pm (x) \to \langle \pi^\pm \rangle + \widetilde{\pi}^\pm (x)/\sqrt{N}$
and normalize the photon field as $A_\mu \to \sqrt{N/3} \cdot A_\mu$.
Taking the large $N$ limit, i.e. $1/N \ll e_0$,
we obtain 
\begin{eqnarray}
i \int d^4 x ({\cal L}_{\rm ph} + \Delta {\cal L} ) 
  - iN {\rm tr} J^{\rm ph}_2
&\simeq& \frac{iN}{6} \int d^4 x A_\mu
  \biggl[ \left( g^{\mu\nu} \partial^2 - \partial^\mu \partial^\nu \right) 
     \left( 1 + N \Pi^{\rm R}(x) \right)
\nonumber \\
&&+ \frac1{\xi} \partial^\mu \partial^\nu 
  + N g^{\mu\nu} \pi^+ \pi^- 
   \Pi_{\pi\pi}^{\rm R} (x) \biggr] A_\nu ,
\label{l2_i2_l}
\end{eqnarray}
where the renormalized self-energies read
\begin{equation}
\Pi^{\rm R} (p^2) \equiv
  \Pi (p^2) - \Pi (0)
 = \frac{5 \alpha}{27 \pi}  
   \left[ \frac{8}{3} - h^2
   - \frac{h}{2}(3-h^2) 
   \ln \frac{h+1}{h-1} \right]
\label{pi_r}
\end{equation}
and
\begin{equation}
\Pi_{\pi\pi}^{\rm R} (p^2) \equiv
  \Pi_{\pi\pi} (p^2) - \Pi_{\pi\pi} (0)
 = \frac{2 \alpha}{\pi}
   \left( 2
   - h \ln \frac{h+1}{h-1} \right) .
\label{pi_p_r}
\end{equation}

Performing the path integral over $A_\mu$ and the Wick rotation, 
$t\rightarrow -ix_4$, we obtain
\begin{eqnarray}
\int &{\cal D} A&  
  \exp \left[i \int d^4 x ({\cal L}_{\rm ph} + \Delta {\cal L} ) 
  - iN {\rm tr} J^{\rm ph}_2 \right] \nonumber \\
  &=& \exp \left[ \frac{iN}{6} {\rm tr}
      \left\{ \ln(\delta_{\mu\nu} 
      + N \pi^+ \pi^- \Pi_{\pi\pi}^{\rm R} D_{\mu\nu})
       - \ln D_{\mu\nu} \right\} \right] ,
\label{za_1}
\end{eqnarray}
where {\rm tr} represents the trace operation over the space-time coordinates
and $D_{\mu\nu}$ represents the photon propagator
\begin{equation}
D_{\mu\nu} (k) 
= \frac1{k^2 (1+N\Pi^R)}
    \left( \delta_{\mu\nu} - \frac{k_\mu k_\nu}{k^2} \right)
  + \xi \frac{k_\mu k_\nu}{k^4} .
\end{equation}
Substituting Eqs. (\ref{pi_r}) and (\ref{pi_p_r}) into Eq.~(\ref{za_1}),
we obtain the generating functional in the Landau gauge,
\begin{eqnarray}
Z &\simeq& Z_{\rm m} 
 + \int {\cal D} \sigma {\cal D} \pi \exp \left[ \frac{iN}{6} 
   \int d^4 x \int \frac{d^4 p}{(2\pi)^4} \right. \nonumber \\
&&\times \left\{ 3\ln \left[ p^2 + 4\sigma'^2
     \left( 1+\frac{4\alpha N}{27\pi}  \right)
  - \frac{4\alpha N}{3\pi} \pi^+ \pi^- \right] \right. \nonumber \\
&&+ \left. \left. \ln \left[ p^2 + 4\sigma'^2 
     \left( 1+\frac{4\alpha N}{27\pi}  \right) \right]
  - 4\ln (p^2 + 4\sigma'^2)
    \right\} \right] . 
\label{za_2}
\end{eqnarray}
In Eq.~(\ref{za_2}),
we omit terms which do not depend on $\sigma$ and $\pi$ and apply the
following approximation:
\begin{eqnarray}
h(p^2) \ln \frac{h(p^2) +1}{h(p^2) -1}
= 2 \left[ 1 + \frac{1}{3 h(p^2)^2} + \frac{1}{5 h(p^2)^4}
  + \cdots \right] 
\quad {\rm for}\ p^2 \sim 0 . 
\label{app_1}
\end{eqnarray}
After the momentum integration,
the effective potential is found to be 
\begin{eqnarray}
V(\sigma,\pi^a)
 = V_{\rm m}(\sigma,\pi^a) + V_{\rm gau}(\sigma,\pi^a) ,
\label{v_a_old}
\end{eqnarray}
where $V_{\rm gau}(\sigma,\pi^a)$ is given by
\begin{eqnarray}
V_{\rm gau}(\sigma,\pi^a)
&=& \frac{1}{192\pi^2} \left[ 
 3f \left( 4\sigma'^2 ( 1 + 4\alpha N / 27\pi) 
 - (4\alpha N / 3\pi) \pi^+ \pi^- 
  ; \Lambda_{\rm p}{}^2 \right) \right.
\nonumber \\
&& \left. 
 + f \left( 4\sigma'^2 ( 1 + 4\alpha N / 27\pi) 
  ; \Lambda_{\rm p}{}^2 \right)
 - 4f(4\sigma'^2; \Lambda_{\rm p}{}^2) \right].
\end{eqnarray}
Here,
the function $f(s^2; t^2)$ is defined in Eq.~(\ref{fst}), 
and $\Lambda_{\rm f}$ and $\Lambda_{\rm p}$ are the three-momentum 
UV cutoffs for the quark and photon momenta, respectively.

\begin{figure}[t]
 \begin{center}
 \begin{tabular}{cc}
  \includegraphics[width=0.45\textwidth]{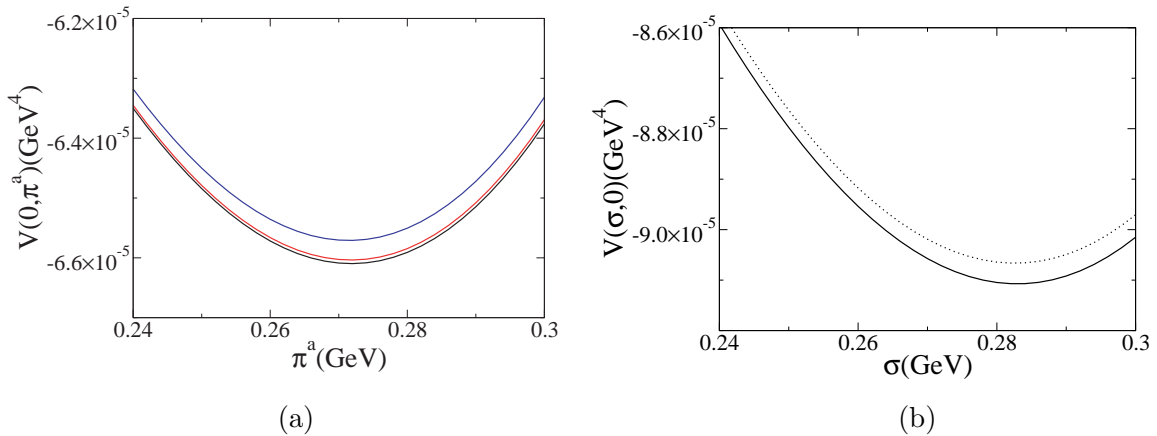}
  &
  \includegraphics[width=0.45\textwidth]{fig4b.eps}
  \\ \quad
   {\small (a)}
  & \quad \quad \quad
   {\small (b)}
  \end{tabular}
 \end{center}
\caption{
 Behavior of the effective potential (\ref{v_a_old}) for
 $m_u+m_d=9$ MeV.
In (a), the black curve represents $V(0,\pi^a)$ at $e= 0$.
The red and blue curves represent $V(0,\pi^{1,2},0)$ and $V(0,0,\pi^3)$, respectively.
In (b), the solid and dotted curves represent $V(\sigma,0,0)$ at
 $e= 0$ and $V(\sigma,0,0)$, respectively.}
\label{v_m45_old}

\end{figure}

To study the QED corrections for the phase structure, we numerically
evaluated the effective potential (\ref{v_a_old}).
In these calculations, we set $\Lambda_{\rm f} \equiv \Lambda_{\rm p}$ and chose the average
of the current quark mass to be $m = 4.5$ MeV. The behavior of the effective 
potential is shown in Fig. \ref{v_m45_old}.
It is seen that the QED correction slightly suppresses
the symmetry breaking for $V(\sigma,\pi^a=0)$ and
$V(\sigma=0, \pi^{1,2}=0, \pi^3)$
but enhances it for $V(\sigma=0,\pi^{1,2}, \pi^3=0)$.
This result for $\sigma$ and $\pi^3$ is
different from these obtained in previous studies\cite{Mira_89}.
This result depends on the procedure used to regularize the theory.
However, the global minimum of the effective potential lies
on the line $\langle \pi^a \rangle =0$. Only the scalar composite field
$\sigma$ acquires a non-vanishing vacuum expectation value, and the 
orientation of the chiral symmetry breaking remains in
the $\sigma$ direction even with the QED correction at the one loop
level. 
Taking the massless limit, $m \to 0$, the effective potential in the 
$\sigma$ direction coincides with that in the $\pi^3$ direction.

Because of the QED correction,
the second derivative of the effective potential in terms of $\pi^{1,2}$
becomes smaller than that of $\pi^3$.
Thus, to leading order in the 1/N expansion, we obtain
a result for the charged and the neutral pion mass
difference whose sign is the opposite of the observed
value.
We consider this further in the next section.

\section{Charged and neutral pion mass}
As is well-known, a pion-photon interaction plays
an essential role in determining the mass difference
between charged and neutral pions\cite{Dmitrasinovic:1992hb}.
Such an interaction is not included in the NJL model
to leading order in the $1/N$ expansion.
Here, we phenomenologically introduce 
higher-order corrections through the kinetic term for mesons
\cite{Suga_91},
\begin{equation}
{\cal L}_{\rm s} = \frac1{2N} \left[ 
 \frac12 (\partial_\mu \sigma)^2
  + \frac12 (\partial_\mu \pi^0)^2
  + (\partial_\mu + ieA_\mu) \pi^+
    (\partial_\mu - ieA_\mu) \pi^- \right]  ,
\label{lag_s}
\end{equation}
where we fix $N=3$, because this term appears at the next-to-leading order
in the $1/N$ expansion. The total Lagrangian is given by
\begin{equation}
{\cal L} = {\cal L}_{\rm f} + {\cal L}_{\rm ph} + {\cal L}_{\rm s} .
\label{lag_fsp} 
\end{equation}

\begin{figure}[t] 
 \begin{center}
  \includegraphics[width=100mm,clip]{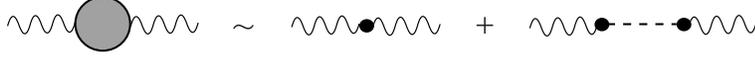}
    \caption{ \small{Leading-order diagrams of the vacuum polarization for photons.
      The wavy lines and the dashed line represent the photon and
      charged pion propagators, respectively.}
    }
  \label{phot}
 \end{center}
\end{figure}

The expectation value of the charged pions induces
the photon mass  \\
$m_{\rm A}{}^2\equiv e^2\pi^+\pi^-/3$.
This seems to suppress the QED correction for
the effective potential $V(\sigma=0,\pi^{1,2},\pi^3=0)$.

To determine it, we calculate the effective potential for
the theory with the Lagrangian (\ref{lag_fsp}). 
The leading-order corrections to the vacuum polarization
amplitude are represented by diagrams in Fig.~\ref{phot}.  
Only the charged pions contribute to these corrections.
As shown in Refs. \citen{Gura} and \citen{Dash_69},
the neutral pion mass should be invariant under
the QED correction.
 After the normalization
$A_\mu \to \sqrt{N/3} A_\mu$, we define the effective action by

\begin{eqnarray}
\int {\cal D} A 
 \exp \left[ i \int ( {\cal L}_{\rm ph} + {\cal L}_{\rm s} 
   + \Delta{\cal L}) - iN {\rm tr} J^{\rm ph}_2 \right]
= \exp \left\{ iN \Gamma[\sigma, \pi^a] \right\} .
\end{eqnarray} 
To leading order in the $1/N$ expansion,
it reads
\begin{eqnarray} 
\Gamma[\sigma,\pi^a] 
&\simeq& \frac{i}{6} \ln \det
 \left[ (- g^{\mu\nu} \partial^2 + \partial^\mu \partial^\nu ) 
     ( 1 + N \Pi^{\rm R} ) - \frac1{\xi} \partial^\mu \partial^\nu
 \right.
\nonumber \\
&&\left. 
      - N g^{\mu\nu} \pi^+ \pi^- \Pi_{\pi\pi}^{\rm R} 
      - m_{\rm A}{}^2 \left( g^{\mu\nu} 
      - \frac{\partial^\mu \partial^\nu}{\partial^2} \right)
  \right] .
\label{gam_ph}
\end{eqnarray}
In the Landau gauge
the effective potential is given by
\begin{eqnarray}
V_{\rm gau}(\sigma,\pi^a) 
&\equiv&
  - \frac1{2\int d^4 x} \Gamma[\sigma,\pi^a] \nonumber \\
&=& \frac1{12} \int \frac{d^4 p}{i(2\pi)^4} 
  \left\{ 3\ln \left[ 4\sigma'^2 \left( 1+\frac{4\alpha N}{27\pi} \right)
    - \frac{4\alpha N}{3\pi} \pi^+ \pi^- - p^2 \right] 
  \right. \nonumber \\
&&+\left. \ln \left[ 4\sigma'^2 \left( 1+\frac{4\alpha N}{27\pi}  \right) 
    - p^2 \right]
    - 4\ln (4\sigma'^2 - p^2)
         + 3\ln (m_{\rm A}{}^2 - p^2) \right\} .
\nonumber \\
\end{eqnarray}
Thus the total effective potential can be written 
\begin{eqnarray}
V(\sigma,\pi^a) = V_{\rm m} (\sigma,\pi^a) 
 + V_{\rm gau+pi} (\sigma,\pi^a) ,
\label{v_a}
\end{eqnarray}
where $V_{\rm gau+pi} (\sigma,\pi^a)$ is given by 
\begin{eqnarray}
V_{\rm gau+pi}(\sigma,\pi^a) = V_{\rm gau} (\sigma,\pi^a) 
 + \frac1{64\pi^2} f(m_{\rm A}{}^2;\Lambda_{\rm p}^2) .
\end{eqnarray}

\begin{figure}[t]
 \begin{center}
  \includegraphics[width=70mm,clip]{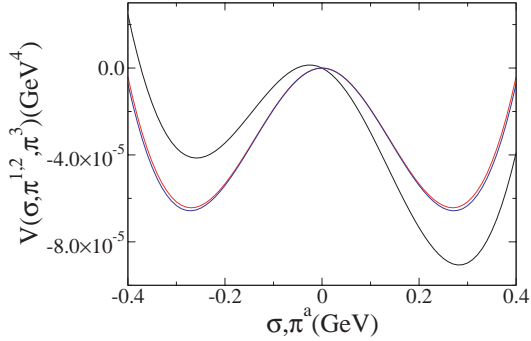}
    \caption{\small Behavior of the effective potential (\ref{v_a}) 
      for $m_{\rm u}+m_{\rm d}=9$ MeV. 
      The black curve represents $V(\sigma, \pi^a=0)$, 
      and the red and blue curves represent
      $V(\sigma=0, \pi^{1,2}, \pi^3=0)$ and 
      $V(\sigma=0, \pi^{1,2}=0, \pi^3)$, respectively.
     }
  \label{v_m45}
 \end{center}
\end{figure}

We numerically analyzed the effective potential for 
$\Lambda_{\rm f} \equiv \Lambda_{\rm p}$ at $m = 4.5$ MeV in this case also.
The resulting behavior is plotted 
along the $\sigma$, $\pi^{1,2}$ and $\pi^3$ axes
in Fig.~\ref{v_m45}.
Comparing these results with those in Fig.~\ref{v_m45_old},
we find that
the effective potential along the $\pi^{1,2}$ axis,
$V(\sigma=0,\pi^{1,2},\pi^3=0)$, is higher.
The curvature of the effective potential increases along
the $\pi^{1,2}$ direction around the minimum.
Thus the charged pion acquires
the larger mass through the pion-photon interaction.
Inserting Eq.~(\ref{v_a}) into Eq.~(\ref{mas_dif}), the mass difference
between the charged and neutral pions is found to be 
\begin{equation}
m_{\rm \pi^{\pm}} - m_{\rm \pi^0}
\simeq 7.9 {\rm~MeV} .
\end{equation}
This value is larger than the observed one.
We note that this result depends on the NJL parameters
$\Lambda_f$ and $\Lambda_p$.
This implies a non-negligible uncertainty in our result.
 For this reason,
if we consider phenomena in which the isospin breaking of the pion mass
may contribute,
we must include higher-order corrections in the $1/N$ expansion.

\section{Gauged NJL model at finite temperature}

It is conjectured that the order parameters of the symmetry
breaking $\langle \sigma \rangle$ and $\langle \pi^a \rangle$
disappear at the critical temperature in the chiral limit, $M\to 0$.
The broken chiral symmetry is restored at higher temperature,
and a state of deconfined partonic matter is realized.
Here we consider a thermal system below the temperature
at which the expectation values $\langle \sigma \rangle$ 
and $\langle \pi^a \rangle$ disappear.

Following the standard procedure of the imaginary time formalism
\cite{Ezaw, Abri}, we introduce the temperature into our model.
We then obtain the effective potential
for the NJL model (\ref{lag}) 
to leading order in the $1/N$ expansion
at finite temperature as \cite{IKM}
\begin{eqnarray}
V_{\rm m}^\beta(\sigma,\pi^a) 
&=& \frac{1}{4G} \left[ \sigma^2 + (\pi^a)^2 \right]
 - \frac{2}{\beta} \sum_{n=-\infty}^\infty 
  \int \frac{d^3 {\boldsymbol k}}{(2 \pi)^3} 
  \ln \left\{(\omega_n^{\rm F})^2 + {\boldsymbol k}^2 + \sigma'^2 \right\} 
\nonumber \\
&& - \frac{2}{\beta} (m_{\rm u} + m_{\rm d}) \sigma
  \sum_{n=-\infty}^\infty 
  \int \frac{d^3 {\boldsymbol k}}{(2 \pi)^3}
  \frac{1}{(\omega_n^{\rm F})^2 + {\boldsymbol k}^2 + \sigma'^2}
\nonumber \\
&& - \frac{1}{\beta} \sum_{n=-\infty}^\infty 
  \int \frac{d^3 {\boldsymbol k}}{(2 \pi)^3}
  \frac{1}{\{(\omega_n^{\rm F})^2 + {\boldsymbol k}^2 + \sigma'^2\}^2}
\nonumber \\
&& \times \left[ (m_{\rm u}{}^2 + m_{\rm d}{}^2)
      \left\{(\omega_n^{\rm F})^2 + {\boldsymbol k}^2 + \sigma'^2 \right\} 
   \right.
\nonumber \\
&& \left.
   - 2 \left\{ (m_{\rm u}{}^2 + m_{\rm d}{}^2) \sigma^2
    + (m_{\rm u} - m_{\rm d})^2 \pi^+ \pi^- \right\} \right] ,
\label{epotmb}
\end{eqnarray}
where $\beta$ is the inverse temperature, $\beta=1/T$, and
the discrete variable $\omega_n^{\rm F}$ is given by
$\omega_n^{\rm F} = (\pi/\beta) (2n + 1)$,
with anti-periodic boundary conditions for fermions.
Using the formulae
\begin{eqnarray}
\sum_{n=-\infty}^\infty
 \frac{1}{(\omega_n^{\rm F})^2 + x^2}
&=& \frac{\beta}{2x} \tanh \frac{\beta}{2} x ,
\label{fml1}
\\
\sum_{n=-\infty}^\infty \ln \left\{(\omega_n^{\rm F})^2 + x^2 \right\}
&=& 2 \ln \cosh \frac{\beta}{2} x
    + {\rm [constant]} ,
\label{fml2} 
\end{eqnarray}
we carry out the summation in Eq. (\ref{epotmb}) and get
\begin{eqnarray}
V_{\rm m}^\beta(\sigma,\pi^a)
&=& \frac{1}{4G} \left[ \sigma^2 + (\pi^a)^2 \right]
  - \frac{2}{\pi^2 \beta} \int_0^{\Lambda_{\rm f}} dk\ k^2
  \ln \cosh \frac{\beta}{2} \sqrt{k^2 + \sigma'^2}
\nonumber \\
&& - \frac{1}{2 \pi^2} (m_{\rm u} + m_{\rm d}) \sigma 
  \int_0^{\Lambda_{\rm f}} dk \frac{k^2}{\sqrt{k^2 + \sigma'^2}}
  \tanh \frac{\beta}{2} \sqrt{k^2 + \sigma'^2}
\nonumber \\
&& - \frac{1}{4 \pi^2} (m_{\rm u}{}^2 + m_{\rm d}{}^2) 
  \int_0^{\Lambda_{\rm f}} dk \frac{k^2}{\sqrt{k^2 + \sigma'^2}}
  \tanh \frac{\beta}{2} \sqrt{k^2 + \sigma'^2}
\nonumber \\
&& + \frac{1}{4 \pi^2} \left\{ (m_{\rm u}{}^2 + m_{\rm d}{}^2) \sigma^2
     + (m_{\rm u} - m_{\rm d})^2 \pi^+ \pi^- \right\}
\nonumber \\
&& \times \int_0^{\Lambda_{\rm f}} dk \frac{k^2}{k^2 + \sigma'^2}
   \left[ \frac{1}{\sqrt{k^2 + \sigma'^2}}
     \tanh \frac{\beta}{2} \sqrt{k^2 + \sigma'^2} \right.
\nonumber \\
&& 
   \left. - \frac{\beta}{2} 
     {\rm sech}^2 \frac{\beta}{2} \sqrt{k^2 + \sigma'^2} \right] .
\label{v_mt}
\end{eqnarray}

\begin{figure}[t]
 \begin{center}
  \includegraphics[width=70mm,clip]{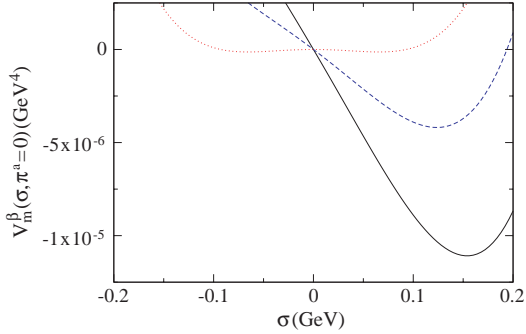}
    \caption{ \small Behavior of the effective potential $(\ref{v_mt})$.
      The red curve represents $V^\beta_{\rm m}(\sigma, \pi^a=0)$ for 
      $m_{\rm u}+m_{\rm d}=0$ MeV, 
      the blue curve represents $V^\beta_{\rm m}(\sigma, \pi^a=0)$ for 
      $m_{\rm u}+m_{\rm d}=4$ MeV, and 
      the black curve represents $V^\beta_{\rm m}(\sigma, \pi^a=0)$ for
      $m_{\rm u}+m_{\rm d}=9$ MeV at $T=170$ MeV. }
  \label{sig_mt}
 \end{center}
\end{figure}

For $m_{\rm u}=2.5$ MeV and $m_{\rm d}=6.5$ MeV, the global minimum of 
the effective potential lies in the $\pi^a=0$ plane.
In Fig. \ref{sig_mt}, we plot the behavior of the effective 
potential $V^\beta_{\rm m}(\sigma, \pi^a=0)$ in this plane.
In the ground state, only the scalar field $\sigma$
develops a non-vanishing expectation value.
As seen in Fig. \ref{sig_mt}, a small mass difference between
up and down quarks induces a larger difference for the effective 
potential near the critical temperature.
Since the effective potential in the $\sigma=0$ plane,
$V_{\rm m}^\beta(\sigma=0, \pi^a)$,
has only a weak dependence on the current quark masses,
$m_{\rm u}$ and $m_{\rm d}$,
$V_{\rm m}^\beta(\sigma=0, \pi^a)$ for
$m_{\rm u} + m_{\rm d}$ on the order of a few MeV has a shape
that is similar
to that for $m_{\rm u} + m_{\rm d} = 0$.
In the limit $m_{\rm u} + m_{\rm d} \rightarrow 0$, isospin symmetry
breaking appears in terms of $O(m^2)$.
Thus the shape of the effective potential $V_{\rm m}^\beta(\sigma=0, \pi^a)$ is
almost the same as that of $V_{\rm m}^\beta(\sigma, \pi^a=0)$,
which is plotted by the red curve in Fig. \ref{sig_mt}.
The difference between the shapes of the effective potentials
in the $\pi^a=0$ plane and the $\sigma=0$ plane is extremely small.
In fact, we cannot distinguish these shapes 
in Fig. \ref{sig_mt}.

It should be noted that the isospin symmetry breaking arises 
from the difference between the masses $m_{\rm u}=2.5$ MeV
and $m_{\rm d}=6.5$ MeV. 
This causes the small differences between 
$\langle \pi^{1,2} \rangle_{\sigma=0, \pi^3=0}$
and $\langle \pi^3 \rangle_{\sigma=0, \pi^{1,2}=0}$.
The ratio for the local minimum,
\begin{equation}
r_\pi \equiv
\left| \frac{\langle \pi^{1,2} \rangle_{\sigma=0, \pi^3=0}
            - \langle \pi^{3} \rangle_{\sigma=0, \pi^{1,2}=0}}
            {\langle \pi^{1,2} \rangle_{\sigma=0, \pi^3=0}
            + \langle \pi^{3} \rangle_{\sigma=0, \pi^{1,2}=0}}
\right| ,
\end{equation}
is approximately $0.05 \%$ for $m_{\rm u}=2.5$ MeV
and $m_{\rm d}=6.5$ MeV at $T = 170$ MeV.
The effective potential at the point
$\sigma=0, \pi^{1,2}=0, \pi^3=\langle \pi^{3} \rangle_{\sigma=0, \pi^{1,2}=0}$
is slightly smaller than that at
$\sigma=0, \pi^{1,2}=\langle \pi^{1,2} \rangle_{\sigma=0, \pi^3=0}, \pi^{3} =0$.
The ratio for the depth of the effective potential,
$r_{\rm V}$ given in (\ref{rv}), 
is approximately $0.2 \%$ at $T = 170$ MeV.
In the case $m_{\rm u}+m_{\rm d}\lesssim 0.0038$ MeV, we again observe  
neutral pion condensation.

In the gauged NJL model, the situation is somewhat complicated.
Due to the Debye screening photon mass
$m_\beta{}^2 = (5/54)e^2 T^2 + O(T)$, 
the photon propagator is modified at finite temperature \cite{LeBe}. 
In the case of electrostatic shielding, $p_0 \to 0$, it is expressed as
\begin{eqnarray}
D_{\mu\nu} &=& \frac{1}{p^2}P_{\mu\nu}^{\rm T}
 + \frac{1}{p^2+2m_\beta{}^2} P_{\mu\nu}^{\rm L}
 + \frac{\xi}{p^2} \frac{p_{\mu}p_{\nu}}{p^2} , 
\label{d_t}\\
P_{44}^{\rm T} &\equiv& 0, \quad
 P_{4j}^{\rm T} \equiv  0, \quad 
 P_{ij}^{\rm T} \equiv \delta_{ij} 
   - \frac{p_i p_j}{{\boldsymbol p}^2} , \\
P_{\mu\nu}^{\rm L} &\equiv& \delta_{\mu\nu} - \frac{p_\mu p_\nu}{p{}^2}
 - P_{\mu\nu}^{\rm T} .
\end{eqnarray}

Replacing the first two terms of Eq.~(\ref{gam_ph}) with
Eq.~(\ref{d_t}), we obtain the QED corrections to
the effective potential at finite
temperature in the Landau gauge,
\begin{eqnarray} 
V_{\rm gau+pi}^\beta (\sigma,\pi^a) 
&=& \frac1{12 \beta} \sum_{n=-\infty}^\infty 
  \int \frac{d^3 {\boldsymbol p}}{(2\pi)^3}
  \left[ 3\ln \left\{ (\omega_n^{\rm B})^2 + {\boldsymbol p}^2
    + 4\sigma'^2 - \frac{4\alpha N}{3\pi} \pi^+\pi^-  \right\}
  \right. \nonumber \\
&&- 3\ln \left\{ (\omega_n^{\rm B})^2 + {\boldsymbol p}^2
    + 4\sigma'^2 \right\}
  + 2\ln \left\{ (\omega_n^{\rm B})^2 + {\boldsymbol p}^2 
     + m_{\rm A}{}^2 \right\}
\nonumber \\
&&\left. + \ln \left\{(\omega_n^{\rm B})^2 + {\boldsymbol p}^2 
     + m_{\rm A}{}^2 + 2m_\beta{}^2 \right\}
  \right] ,
\end{eqnarray}
where the square of the four momentum $p^2$ is written 
$p^2=(\omega^{\rm B}_n)^2+{\boldsymbol p}^2$, and the discrete variable 
$\omega^{\rm B}_n$ is given by 
$\omega_n^{\rm B}= (\pi/\beta) 2n$,
from the periodic boundary conditions for bosons.

From the above,
 the effective potential of the gauged NJL model
is found to be
\begin{equation}
  V^\beta(\sigma,\pi^a) = 
   V_{\rm m}^\beta(\sigma,\pi^a) + V_{\rm gau+pi}^\beta (\sigma,\pi^a) .
\label{v_t}
\end{equation}
The photon dependent part $V_{\rm gau+pi}^\beta (\sigma,\pi^a)$
is given by
\begin{eqnarray}
V_{\rm gau+pi}^\beta (\sigma,\pi^a)
&=& \frac{1}{12\pi^2\beta} \int_0^{\Lambda_{\rm p}} dp\ p^2
   \left[ 3\ln \sinh \frac{\beta}2
      \sqrt{ p^2 + 4\sigma'^2 - \frac{4\alpha N}{3\pi} \pi^+\pi^- } 
   \right.
\nonumber \\
&&- 3\ln \sinh \frac{\beta}2 \sqrt{p^2 + 4\sigma'^2}
  + 2\ln \sinh \frac{\beta}2 \sqrt{p^2 + m_{\rm A}^2} 
\nonumber \\
&& \left. + \ln \sinh \frac{\beta}2
     \sqrt{p^2 + m_{\rm A}^2 + 2m_\beta{}^2} \right] ,
\label{v_at}
\end{eqnarray}
where we have used the formula
\begin{equation}
\sum_{n=-\infty}^\infty \ln \left\{(\omega_n^{\rm B})^2 + x^2 \right\}
= 2 \ln \sinh \frac{\beta}{2} x  
    + {\rm [constant]} .
\end{equation}

\begin{figure}[t]
 \begin{center}
  \includegraphics[width=65mm,clip]{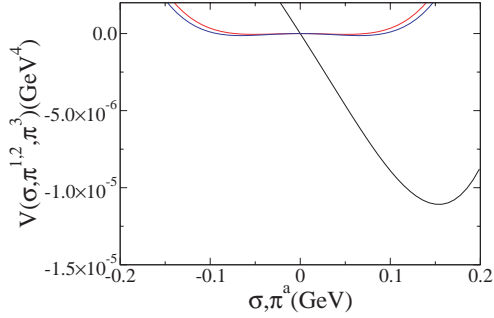}
    \caption{\small Behavior of the effective potential (\ref{v_t}) for 
      $m_{\rm u}+m_{\rm d}=9$ MeV at $T=170$ MeV.
      The black curve represents $V^\beta(\sigma, \pi^a=0)$, 
      and the red and blue curves represent 
      $V^\beta(\sigma=0,\pi^{1,2},\pi^3=0)$
      and $V^\beta(\sigma=0, \pi^{1,2}=0,\pi^3)$, respectively. }
  \label{v_m45_t170}
 \end{center}
\end{figure}

We numerically calculated Eq.~(\ref{v_t}) with Eqs.~(\ref{v_mt}) 
and  (\ref{v_at}) just below the critical temperature.
In Fig. \ref{v_m45_t170}, we restrict the parameter space
to the region in which
two of the composite fields $\sigma$, $\pi^{1,2}$ and $\pi^3$ 
vanish and plot the effective potential (\ref{v_t}) for $T=170$ MeV. 
For $m_{\rm u}+m_{\rm d}=9$ MeV, we find the global minimum of the 
effective potential at $\pi^{1,2}=\pi^3=0$. Only the 
composite field $\sigma$ condenses even at finite temperature. 
The orientation of the chiral symmetry breaking is again preserved 
under the QED correction. 

\begin{figure}[t]
 \begin{center}
  \includegraphics[width=65mm,clip]{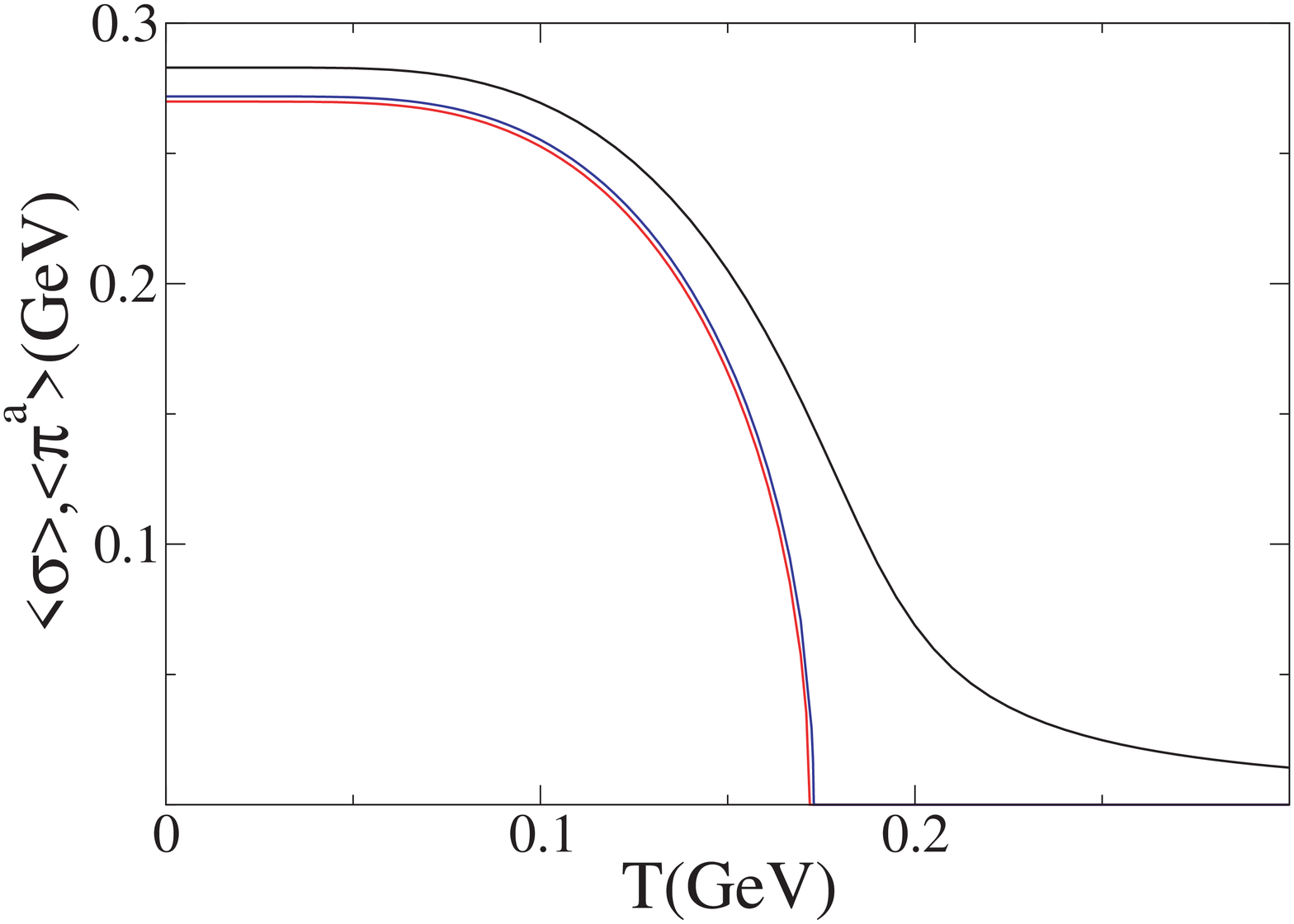}
    \caption{\small Behavior of the global and local minima of 
      the effective potential 
      (\ref{v_t}) for $m_{\rm u}+m_{\rm d}=9$ MeV.
      The black curve represents the location of the global minimum,
      $\langle \sigma\rangle_{\pi^a=0}$. 
      The red and blue curves represent the local minima 
      $\langle \pi^{1,2} \rangle_{\sigma=\pi^3=0}$ 
      and $\langle \pi^3 \rangle_{\sigma=\pi^{1,2}=0}$, respectively.}
  \label{gap_m45}
 \end{center}
\end{figure}

In Fig.~\ref{gap_m45}, we plot the minimum of the effective potential in 
the restricted parameter space, $V^\beta(\sigma, \pi^a=0)$, 
$V^\beta(\sigma=0, \pi^{1,2}, \pi^3=0)$ and 
$V^\beta(\sigma=0, \pi^{1,2}=0, \pi^3)$.
As is clearly seen in the figure, the expectation value 
$\langle \sigma\rangle$ disappears smoothly  at higher temperature.
Thus, the transition from the broken phase to the symmetric phase
is crossover. 
 This is a well-known feature of the theory with a current quark 
mass. It should be noted that the pion fields have vanishing
expectation values in the ground state for $m_{\rm u}+m_{\rm d}=9$MeV.
It is conjectured that the chiral symmetry is restored in the early universe.
Non-vanishing expectation values of the pion fields
$\langle\pi^{1,2}\rangle_{\sigma=0, \pi^3=0}$ and
$\langle\pi^{3}\rangle_{\sigma=0, \pi^{1,2}=0}$
may be realized during the transition period of the chiral symmetry
breaking, i.e. ``disoriented chiral condensation''\cite{Raja, Moha}.

Near the critical temperature we observe that
the ratio $r_{\rm V}$, which is defined in Eq. (\ref{rv}), is much larger 
than in the $T = 0$ case. It is believed that the influence of the
isospin breaking may be observed in phenomena near the critical 
temperature.

\begin{figure}[t]
 \begin{center}
  \begin{tabular}{cc}
   \includegraphics[height=0.18\textheight]{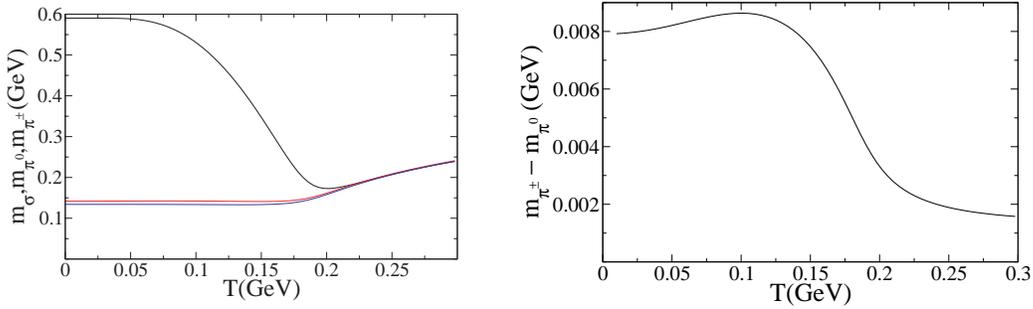}
  &
   \includegraphics[height=0.18\textheight]{fig10b.eps}
  \\ \quad
   {\small (a) Charged and neutral pions masses.}
  & \quad \quad \quad
   {\small (b) Pion mass difference.}
  \end{tabular}
 \end{center}
\caption{Behavior of the pion mass for $m_u+m_d=9\mbox{~MeV}$.
 The red and blue curves represent the charged and neutral pion
 masses, respectively}
\label{mas_pi}
\end{figure}

Inserting Eq.~(\ref{v_t}) into Eq.~(\ref{mas_dif})
and using Eq.~(\ref{z_pi}),
we evaluate the mass for the $\pi$ meson or a soft mode which corresponds to $\pi$.
In Fig.~\ref{mas_pi} we plot the behavior of the pion and sigma masses
and the pion mass difference as functions of temperature.
As seen in Fig. 10(a),
the pion and sigma masses exhibit behavior similar to that found in the  
previous studies Refs. \citen{Hats_94} and \citen{Kuni_89}.
A minimum of the sigma mass is found near the temperature
$T=200 \mbox{~MeV}$.
We call this the critical temperature $T_C$.
Above $T_C$, a pion can exist as a soft mode\cite{Hatsuda:1985eb}.
The QED correction enhances the charged pion mass, even at finite
temperature.
The pion mass difference has a maximum value
below the critical temperature.

\section{Pion condensation and dynamical {\it CP} violation}
In the present section, we consider the more precise structure of the 
pion condensation and discuss the dynamical {\it CP} violation induced
in the four-fermion interaction model. 

The QCD Lagrangian is invariant under the {\it CP} transformation except 
for quark mixing terms. 
Here we consider the NJL model (\ref{lag})
as a low energy effective theory of QCD.
This model does not contain any quark mixing terms.
Thus the Lagrangian density
(\ref{lag}) is {\it CP}-invariant.
Because a pseudo-scalar operator is {\it CP} odd, the expectation value of 
the pion field transforms under {\it CP} as
\begin{equation}
 \left\langle \pi^a \right\rangle \simeq
  - \left\langle \frac{G}{N} \bar\psi i \gamma_5 \tau^a \psi \right\rangle
  \rightarrow
  -\left\langle \pi^a \right\rangle \simeq
  \left\langle \frac{G}{N} \bar\psi i \gamma_5 \tau^a \psi \right\rangle .
\end{equation}
Thus, the non-vanishing expectation value of the pion field
spontaneously violates the CP invariance. Such a mechanism for
violating the {\it CP} invariance was suggested long ago 
by Dashen\cite{Dash_71}. A simple four-fermion model to realize 
such a mechanism was proposed in other contexts in 
Refs. \citen{Hashimoto:1992hh} and \citen{Inagaki:1994bv}.

\begin{table}[b]
\begin{center}
\caption{The depth of the effective potential and the ratio $r_a$ for 
$m_{\rm u}+m_{\rm d}=0$.}
\begin{tabular}{|r|r|r|r|r|r|} \hline
  {\footnotesize $T$}
 & {\footnotesize $V(\langle \sigma \rangle, 0, 0)$}
 & {\footnotesize $V(0, \langle \pi^{1,2} \rangle, 0)$}
 & {\footnotesize $V(0, 0, \langle \pi^3 \rangle)$}
 & {\footnotesize $r_{1,2}$}
 & {\footnotesize $r_3$}
\\ 
 {\footnotesize (MeV)}
 & {\footnotesize (GeV$^4$)}
 & {\footnotesize (GeV$^4$)}
 & {\footnotesize (GeV$^4$)}
 &  & 
\\ \hline
  {\footnotesize $0$}
 & {\footnotesize $-6.60 \times 10^{-5}$}
 & {\footnotesize $-6.60 \times 10^{-5}$}
 & {\footnotesize $-6.61 \times 10^{-5}$}
 & {\footnotesize $0$}
 & {\footnotesize $4.24 \times 10^{-4}$}
\\ \hline
  {\footnotesize $170$}
 & {\footnotesize $-1.30 \times 10^{-7}$}
 & {\footnotesize $-1.30 \times 10^{-7}$}
 & {\footnotesize $-1.33 \times 10^{-7}$}
 & {\footnotesize $0$}
 & {\footnotesize $0.00943$}
\\ \hline
\end{tabular}
\end{center}
\end{table}

As shown in the previous sections, pion condensation is observed
for a vanishing average of the current quark mass, $m_{\rm u}+m_{\rm d}=0$.
In Table III we list the values of the effective potential at the minima on
the $\sigma$, $\pi^{1,2}$ and $\pi^{3}$ axes for $m_{\rm u}+m_{\rm d}=0$
at $T=0$ and $T=170$ MeV.
The global minimum lies on the $\langle\pi^{3}\rangle$
axis. 
The temperature enhances the ratio $r_a$ defined by
\begin{equation}
r_a \equiv \frac{V(\langle \sigma \rangle = 0, \langle \pi^a \rangle) 
                  - V(\langle \sigma \rangle, \langle\pi^a \rangle = 0)}
           { V(\langle \sigma \rangle = 0, \langle\pi^a \rangle)
                  + V(\langle \sigma \rangle, \langle\pi^a \rangle=0)} .
\end{equation} 
Hence we see that a non-trivial {\it CP} violating phase is realized for
the NJL model.

\begin{figure}[t]
 \begin{center}
   \resizebox{!}{4.5cm}{\includegraphics{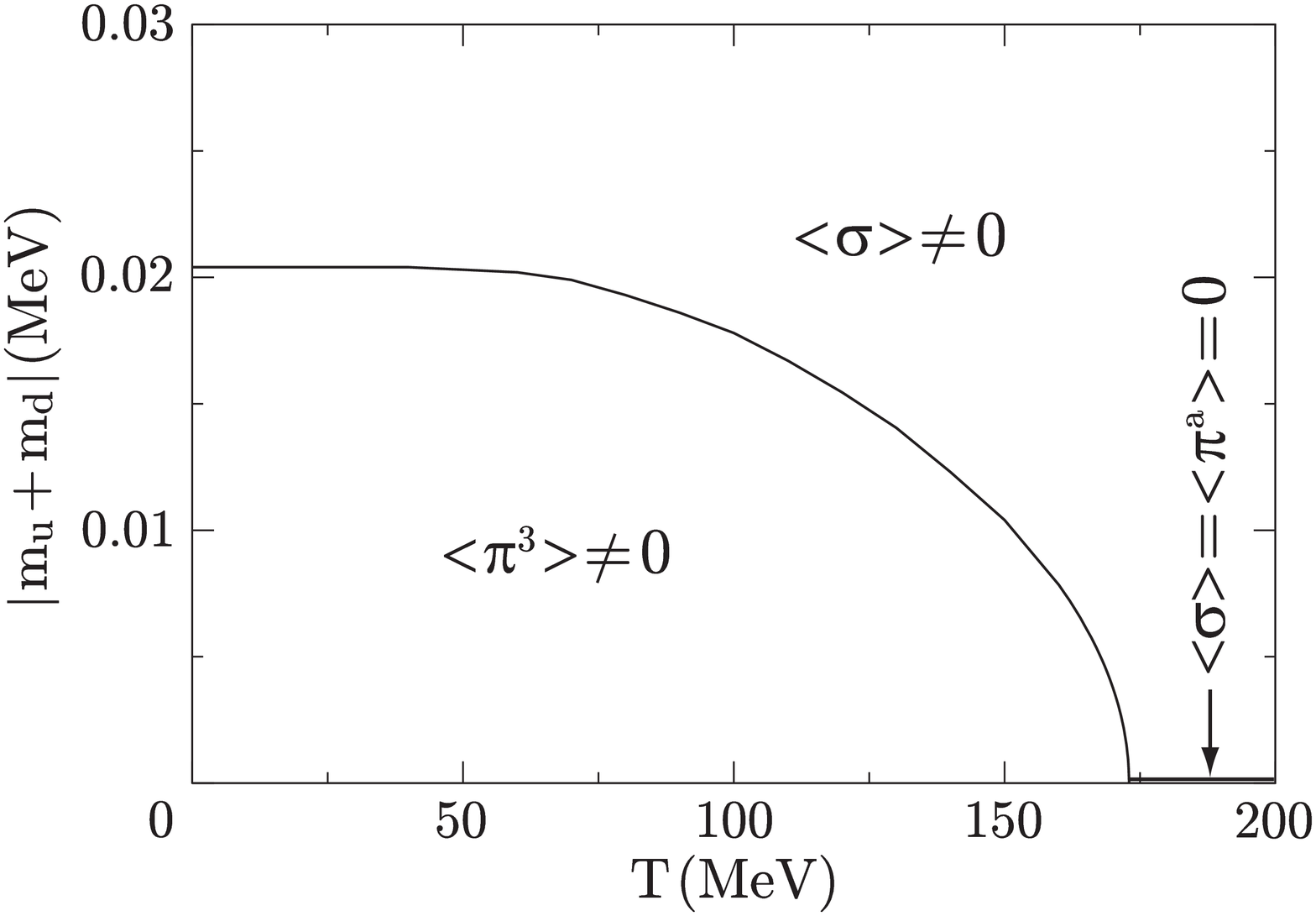}}
 \end{center}
\caption{Phase structure in the $m$-$T$ plane.}
\label{m_t_pln2}
\end{figure}

We also evaluated the effective potential for small but finite values of 
$|m_{\rm u}+m_{\rm d}|~(=2m)$ and obtained the phase structure
of the chiral symmetry breaking in the $m-T$ plane. As seen
in Fig.~\ref{m_t_pln2}, the {\it CP} violating phase, 
$\langle\pi^{3}\rangle\neq 0$,
appears within a narrow parameter range.
In Fig. \ref{m_t_pln2} we cannot observe
the restoration of the broken chiral symmetry, except for 
$m_{\rm u}+m_{\rm d}=0$. The expectation value $\langle\sigma\rangle$ 
disappears smoothly at higher temperature,
e.g. in the case depicted in Fig. \ref{gap_m45}.

 Because our original Lagrangian has no $\theta$-term,
the dynamical {\it CP} violation takes place in the NJL model
with two flavors of light quarks. 
The {\it CP} violating phase is observed for 
$|m_{\rm u} + m_{\rm d}|\lesssim 0.020$ MeV  
at $T=0$. Below, we consider two assumptions concerning 
the current quark mass.

With the assumption
$|m_{\rm u} + m_{\rm d}|\lesssim 0.020$ MeV and 
$|m_{\rm u}| + |m_{\rm d}|\gtrsim 0.020$ MeV,
the up and down quark masses have opposite sign.
The sign of the quark mass is reversed by the discrete 
chiral transformation $u\rightarrow \gamma_5 u$.
Thus, the theory with a mass term, 
$|m_{\rm u} + m_{\rm d}|\gtrsim 0.020$ MeV, is obtained through
the chiral transformation. Under this transformation the expectation
value of the pion field vanishes, and therefore the ground state
preserves the {\it CP} invariance. However, the discrete chiral
transformation is anomalous. It produces the {\it CP} violating 
$\theta$-term with the angle $\pi$ in QCD.
This is known as a dynamical mechanism violating {\it CP}
invariance.\cite{Dash_71}
This term is inconsistent with the experimental upper bound for the neutron
electric dipole moment.
We thus conclude that another contribution to the neutron
electric dipole moment, for example contributions
from heavier particles, should be introduced.

With the other assumption, $|m_u+m_d|\lesssim 0.020$ MeV and
$0<|m_{\rm u}| + |m_{\rm d}|\lesssim 0.020$ MeV,
we cannot obtain the Lagrangian with
$|m_{\rm u} + m_{\rm d}|\gtrsim 0.020$ MeV
through any redefinition of the quark fields.
In particular, if either  
the up or down quark is massless, any $\theta$-term can be rotated away.
This case has been investigated in Refs. \citen{Creu0} and \citen{Creu}.
With the assumption $|m_{\rm u}| + |m_{\rm d}|\lesssim 0.020$ MeV, 
the model is entirely incapable of accounting for
the real meson properties.
But, nevertheless, it is 
interesting to suggest the possibility that dynamical {\it CP} violation 
occurs at low temperature without introducing the strong {\it CP} problem.

\section{Conclusion}
 
We have investigated the symmetry properties of the NJL model with two
flavors of quarks.
The explicit breaking of the global flavor symmetry
$SU_{\rm L}(2)\otimes$ $SU_{\rm R}(2)$ was introduced from the
current quark mass and the QED interaction. We have considered the 
explicit breaking of this symmetry up to $O(m^2)$ and $O(e^2)$ and
calculated the effective potential to leading order in the 
$1/N$ expansion. 

Evaluating the effective potential, we have determined the precise structure
of the chiral symmetry breaking. The effective potential has terms
proportional to the average of the current quark mass
at $O(m)$.
We found that a positive quark mass enhances the chiral symmetry breaking for a
positive $\sigma$ and suppresses it for a negative $\sigma$.
A negative quark mass has the opposite effect. For a 
realistic quark mass, $m_{\rm u}+m_{\rm d}\simeq 9$ MeV,
it was found that
the ground state is characterized by $\pi^a=0$. Only the scalar composite state 
$\sigma$ develops a non-vanishing expectation value. This situation
does not change when we introduce the one-loop QED corrections and
the finite temperature corrections. Therefore we conclude that the 
orientation of the chiral symmetry breaking is invariant for
a sufficiently large magnitude of the average of the quark mass, 
$|m_{\rm u}+m_{\rm d}|\gtrsim 0.020$ MeV.

The difference between the current quark masses of the up and down 
quarks begins to affect the effective potential at second order,
$O(|m_{\rm u}-m_{\rm d}|^2$). It contributes to the $\pi^3$ direction
and causes isospin breaking for the effective potential, but 
the ratio of 
$V(\langle \pi^{1,2} \rangle)$ to $V(\langle \pi^3 \rangle)$ 
is extremely small, $0.008 \%$ at $T=0$ and $0.2 \%$ for 
$m_{\rm u}=2.5$ MeV and $m_{\rm d}=6.5$ MeV at $T=170$ MeV.

These effects cannot change the ground state, but there 
is a possibility that the state passes through a local minimum of 
the effective potential at the time of the symmetry breaking
in the early universe and the pion fields
thus temporarily have a non-vanishing expectation value. 
Highly excited states are produced near the critical temperature;
these excited states may move from the ground state with positive 
$\langle\sigma\rangle$ to a finite $\langle \pi\rangle$ state.
This phenomenon is called ``disoriented chiral condensation''\cite{Raja, Moha}. 

To explain the charged and neutral pion mass difference, we have to
consider higher-order contributions in the $1/N$ expansion.
Here we phenomenologically introduced a pion kinetic term.
Then we obtained the correct sign for
$m_{\pi^\pm}-m_{\pi^0}$.
The value of $m_{\pi^\pm}-m_{\pi^0}$
increases near the critical temperature.
The neutral pion direction is more stable
than the charged pion direction, even at finite temperature.
It is conjectured that the production 
number of neutral pions is larger than that of charged pions when the 
QCD ground state passes from the quark gluon state to the hadron state.

If we assume that the average of the quark mass is sufficiently small, 
the enhancement of the chiral symmetry breaking in the $\sigma$ 
direction will not be large. We cannot ignore the isospin breaking 
terms that come from the mass difference 
even in the ground state. Neutral pion condensation
can be realized in a restricted parameter range. We can consider
the case in which only $\sigma$ develops a non-vanishing expectation 
value through the chiral $SU(2)$ transformation. Then, the only  
difference is found in the $\theta$-term. Thus the pion condensation
studied here affects only phenomena in which the $U_A(1)$ anomaly contributes.
Such pion condensation may not be realized in the present universe, 
but it would be interesting to study the dynamical origin of {\it CP} violation and 
the critical phenomena in the early universe.
Especially for $|m_{\rm u}| + |m_{\rm d}|\lesssim 0.020$ MeV, we find 
the possibility of dynamical {\it CP} violation.

There is a possibility that the composite operator
$\pi^+ \pi^-$ also develops a non-vanishing expectation value.
In the leading-order analysis of the $1/N$ expansion
we cannot evaluate the condensation $\langle \pi^+\pi^-\rangle$.
Thus, an extension of the model or the analysis is necessary to elucidate such
a phenomenon caused by the pion dynamics\cite{KobayashiKugo}.

\section*{Acknowledgements}

The authors would like to thank T.~Morozumi
for fruitful discussions and correspondences. 
T.~I. also thanks A.~Kvinikhidze for stimulating discussions.

\end{document}